# Domain textures in the fractional quantum Hall effect

Ziyu Liu[1,*], Ursula Wurstbauer[2], Lingjie Du[3], Ken W. West[4], Loren N. Pfeiffer[4], Michael J. Manfra[5,6], Aron Pinczuk[1,7,‡]

[1] *Department of Physics, Columbia University, New York, New York 10027, USA*

[2] *Institute of Physics, University of Münster, Wilhelm-Klemm-Str.10, 48149 Münster, Germany*

[3] *School of Physics, and National Laboratory of Solid State Microstructures, Nanjing University, Nanjing 210093, China*

[4] *Department of Electrical Engineering, Princeton University, Princeton, New Jersey 08544, USA*

[5] *Department of Physics and Astronomy, School of Materials Engineering, and School of Electrical and Computer Engineering, Purdue University, Indiana 47907, USA*

[6] *Microsoft Quantum Lab Purdue, Purdue University, West Lafayette, Indiana 47907, USA*

[7] *Department of Applied Physics and Applied Mathematics, Columbia University, New York, New York 10027, USA*

*zl2577@columbia.edu, ‡ap359@columbia.edu

**Impacts of domain textures on low-lying neutral excitations in the bulk of fractional quantum Hall effect (FQHE) systems are probed by resonant inelastic light scattering. We demonstrate that large domains of quantum fluids support long-wavelength neutral collective excitations with well-defined wave vector (momentum) dispersion that could be interpreted by theories for uniform phases. Access to dispersive low-lying neutral collective modes in large domains of FQHE fluids such as long wavelength magnetorotons at filling factor $v = 1/3$ offer significant experimental access to strong electron correlation physics in the FQHE.**

Current geometrical theories of strongly correlated phases in fractional quantum Hall effect (FQHE) fluids of two-dimensional electron systems (2DES) identify low-lying neutral collective excitations (known as magnetorotons [1]) as chiral gravitons [2–7]. The bulk of FQHE fluids is non-uniform and formation of domain textures [8–11] have significant impact on low-lying collective excitations [12–14]. The impact of bulk domain textures in strongly correlated phases is the subject of increasing attention, with prominent examples in FQHE fluids [15–20] and in high temperature superconductors [21–24]. To investigate the insulating bulk of quantum Hall phases, studies on thermal transport and electronic interference by edge modes are interpreted under bulk-edge correspondence [25–31]. Experimental methods that directly probe low-lying neutral collective excitations in the bulk of FQHE fluids are crucial tools in the quest to understand strong electron correlation physics.

In the second Landau level (SLL) at $v = 7/3$ a seemingly conventional FQHE is actually very distinct from its counterpart at $v = 1/3$ in the lowest Landau level (LLL) [32–38]. Even under a small in-plane magnetic field, appearance of anisotropic longitudinal resistance indicates emergence of domains of nematic phases [39,40] that coexist with domains of the bulk FQHE fluid [41]. Newly discovered collective modes under a small in-plane magnetic field are interpreted as long-wavelength plasmons of nematic domains in the filling factor range $2 < v < 3$ [42]. Studies of the plasmons under a small in-plane magnetic field serve as a probe of the impact of domain textures on low-lying collective excitations.

Access to magnetoroton modes of FQHE fluids and to plasmons of nematic liquids is provided by resonant inelastic light scattering (RILS) methods [13,42]. Here we report that RILS by nematic plasmons at $v = 7/3$ under a small in-plane magnetic field reveals a wide range of nematic domain sizes from around 1 $\mu m$ to characteristic sizes larger than several microns, and establishes FQHE-nematic fluids at $v = 7/3$ as an ideal platform to study impacts of domain textures on low-lying neutral excitations. The domains with dimensions much larger than the inelastic scattering wavelength (about 1 $\mu m$) support low-lying collective excitations with well-defined wave vector (momentum) dispersions. Very sharp dispersive long-wavelength magnetoroton modes have been reported in the LLL at $v = 1/3$ [13]. We surmise that the observed modes, with well-defined wave vectors, are excitations of the FQHE fluid in large domains. Results at $v = 1/3$ reported in previous works are examined to highlight possible experimental insights on long-wavelength magnetoroton excitations that are identified as chiral gravitons under strong electron correlation in recent theory for uniform FQHE phases [4–7].

Characteristic domain sizes are obtained by modelling the RILS intensity of nematic plasmons. While the matrix element depends on plasmon-electron interactions [12], we focus here on the photon frequency dependence and on breakdown of wave vector conservation due to formation of domain textures. We find that intensity is [43]:

$$I \propto \left| \frac{1}{(\omega_I - \omega_1 - i\gamma_1)} \frac{1}{(\omega_o - \omega_2 - i\gamma_2)} \right|^2 \frac{1}{(\omega(q) - \omega(k))^2 + (\Delta\omega/2)^2} \quad (1)$$

$\omega_1$ ($\omega_2$) and $\gamma_1$ ($\gamma_2$) are the energies and broadenings of the incoming (outgoing) resonant channels of optical excitations of the GaAs quantum well that hosts 2DES. When both incoming ($\omega_I$) and outgoing ($\omega_o$) photon energies are in the vicinity of these optical excitations, a large intensity enhancement occurs through double resonant inelastic light scattering (DRILS). $\omega(q)$ is the plasmon energy at wave vector $q$ [43]. Energy conservation in inelastic scattering requires $\omega(q) = \omega_I - \omega_o$. $\Delta\omega$ is the range of the plasmon energies due to breakdown of wave vector conservation.

The inelastic scattering wave vector $\boldsymbol{k}$ is along the in-plane magnetic field [42]. Wave vector conservation ($\boldsymbol{k} = \boldsymbol{q}$) occurs for domain sizes $D \gg \lambda = 2\pi/k$, where $\lambda$ is the inelastic scattering wavelength, $k = |\boldsymbol{k}|$ and $D$ is a characteristic length of nematic domains along the direction of $\boldsymbol{q}$. Wave vector conservation is broken under a finite domain size, so mode wave vectors have an uncertainty $\Delta q = 2\pi/D$ [49]. We have

$$\Delta q = \left(\frac{dq}{d\omega}\right)\bigg|_{q=k} \Delta\omega = 2k\Delta\omega/\omega(k) \quad (2)$$

for nematic plasmons with dispersion $\omega(q)$ [43]. Estimates of values of $D$ can be obtained with Eq. (2) and determinations of $\Delta\omega$ from inelastic light scattering spectra.

At $v = 7/3$ there is a set of DRILS spectra for which breakdown of wave vector conservation has very minor impact. Here estimated nematic domain sizes are $D \approx 5.3\ \mu m \gg \lambda = 1.2\ \mu m$. In the same device, strong impact of breakdown of wave vector conservation in another set of spectra is found for domains that have $D$ comparable to $\lambda$. The optical resonances in large nematic domains involve quasiparticle states close to the Fermi level. This could be evidence showing that large nematic domains are responsible for anisotropic electrical conduction in the FQHE at $\nu = 7/3$ under tilted magnetic fields [41]. Domains of FQHE fluid with $D > \lambda$ and those with $D < \lambda$ should coexist in the non-uniform bulk of FQHE states. Albeit the non-uniformity, low-lying neutral collective excitations in large domains with $D \gg \lambda$ accessed by RILS and DRILS would manifest the fundamental correlation physics in the uniform bulk of FQHE phases in the LLL and the SLL.

The ultraclean 2DES is confined in a 30-nm-wide, symmetric, modulation-doped single GaAs/AlGaAs quantum well. Carrier mobility of the wafer measured by transport is $\mu = 23.9 \times 10^6$ cm$^2$/Vs at 300 mK. The sample is mounted on the cold finger of a $^3$He/$^4$He-dilution refrigerator with windows for direct optical access and inserted in the bore of a 16T superconducting magnet. All measurements were performed at T $\leq$ 45 mK. The electron density under illumination is directly determined in each cool down by RILS measurements of spin waves at $\nu = 3$ (Fig. S2) and yields $n = 2.8 \times 10^{15}$ m$^{-2}$ [43]. The stability of electron density and sample quality against illumination are confirmed by photoluminescence (PL) and RILS measurements under zero magnetic field [43]. Figure 1(a) describes the back-scattering geometry at a small tilt angle $\theta = 20°$. The finite wave vector transfer in back-scattering is $k = |\mathbf{k_I} - \mathbf{k_o}|sin\theta$, where $\mathbf{k_I}$ and $\mathbf{k_o}$ are wave vectors of the incoming and outgoing photons. DRILS spectra are excited with the linearly polarized tunable emission from a Ti:sapphire laser that is finely tuned to match the $\omega_1$ excitation. The incident power density was kept well below $10^{-4}$ W/cm$^2$. The outgoing photons are dispersed by a triple grating spectrometer and recorded by a CCD camera.

Optical excitons contributing to the DRILS matrix element are built from transitions between quasiparticles in conduction states and holes in valence states shown in steps 1 and 3 in Figs. 1(b) and 1(c). These resonant channels are identified from PL and resonant Rayleigh scattering spectra at $\nu = 7/3$ as shown in Fig. S1 [43]. At $\nu = 7/3$ excitons participating in DRILS are extremely sharp, indicating that Landau levels in the bulk of nematic-FQHE fluid system support well-defined states even under non-uniform conditions that must prevail due to coexistence of two distinct phases (the FQHE fluid and the nematic liquid). FQHE fluid at $\nu = 7/3$ is characterized by the determination of low energy magnetoroton excitations, which is suppressed by increasing temperatures or deviation of filling factors (Fig. S3) [43,46].

Two types of DRILS spectra of plasmons are reported. They are defined by the outgoing resonances in steps 3 shown in Figs. 1(b) and 1(c). In Low Outgoing Resonance (LOR) mode the outgoing photon resonates with $L$ transitions that involve states of correlated quantum fluids in the SLL [50]. In High Outgoing Resonance (HOR) mode the outgoing photon resonates with the excitonic $X$ transition involving empty states in the partially populated SLL [50]. Figures 2 and 3 report DRILS results by nematic plasmons at $\nu = 7/3$.

In results for DRILS-LOR shown in Fig. 2(a) there is a sharp nematic plasmon peak that slightly blueshifts with higher incoming photon energies within the small energy range $\Delta\omega_S$. The small blueshift is due to minor impact of breakdown of wave vector conservation. Figure 2(b) plots the intensities of plasmons as a function of outgoing (red dots) or incoming (blue dots) photon energies. The blue dots indicate a clear incoming resonance with the $X$ exciton, and the red dots indicate an outgoing resonance associated with the Fermi level of correlated phases [near the high energy edge of $L$ emission in Fig. 2(b)]. We expect the DRILS matrix element to be prominent when symmetry of the final state in $L$ is connected to $X$ exciton through plasmon-electron interactions. The sharp exciton in the outgoing resonance in LOR provides critical spectroscopic insight into links of large nematic domains with the Fermi level of 2DES. The results reveal interplays of topological order in the FQHE with nematic order that could impact anisotropic transport by edge modes [41].

In LOR, an additional broad continuum with energy range $\Delta\omega_B$ shown in Fig. 2(a) is weakly excited without double resonance. We find that in HOR this continuum of modes is DRILS active and gives rise to a series of plasmon modes within $\Delta\omega_B$ shown in Fig. 3(a). Different energy ranges of double resonant plasmons in LOR and HOR spectra reveal a rich set of domain sizes. As domain sizes increase ($D \gg \lambda$), the wave vector dispersions of elementary excitations would tend to be similar to those of a uniform electron fluid.

Modeling DRILS spectra with Eqs. (1) and (2) yields nematic domain sizes $D_S \approx 5.3$ $\mu m$ in LOR and $D_B < 2.2$ $\mu m$ in HOR [43]. Figure 2(c) displays nematic plasmons generated by DRILS model in LOR. Figures 2(d) and 3(b) show DRILS model fits of plasmon energies and intensities as a function of incoming photon energies in LOR and HOR, which are the two factors directly capturing the resonance evolution of plasmons. The fits successfully reproduce DRILS-LOR results. It supports the interpretation that LOR spectra are from large nematic domains with small impact of breakdown of wave vector conservation and thus represent well-defined bulk-like collective modes. In contrast, the HOR incoming resonance [Fig. 3(b)] has a flatter top with a few outliers. When domain sizes are comparable to the inelastic scattering wavelength $\lambda = 1.2$ $\mu m$, wave vectors are no longer good quantum numbers and modeling of DRILS spectra may have to be modified.

It is highly significant that DRILS-LOR involves larger domains where quantum fluids with a sharp optical transition at the Fermi level are better defined. In contrast, DRILS-HOR involves intermediate states of electron-hole pairs forming $X$ and $X+$ excitons which are associated with empty states in the partially populated SLL. As a result, LOR and HOR effectively probe plasmons in two different ranges of nematic domain sizes: larger domains close to the Fermi level in LOR, and smaller domains producing stronger impact of breakdown of wave vector conservation in HOR [Fig. 1(d)]. FQHE-nematic fluids at $v = 7/3$ with DRILS-active plasmons thus clearly reveal distinct impacts of domain textures depending on their sizes.

Domain textures at other filling factors can also be resolved by DRILS. Figure 4(a) shows strong nematic plasmons of a non-FQHE state at $v = 2.68$. DRILS model reproduces main resonance features, as shown in Fig. 4(b). Compared to $v = 7/3$ in LOR, the broader resonance reveals smaller domain sizes $D \approx 3.5$ $\mu m$, which exerts a larger impact of breakdown of wave vector conservation and is consistent with higher disorder level in the upper half of the SLL.

Large domains of FQHE fluids support neutral collective excitations with well-defined wave vector dispersion that could be observed in RILS and DRILS spectra which access bulk states.

Extremely sharp long-wavelength magnetoroton excitations at $v = 1/3$ were observed [13,51]. Single resonance is considered here due to the lack of a pair of well-understood resonant channels with energy difference matching the magnetoroton energy. The narrow linewidth can be regarded as a consequence of very small breakdown of wave vector conservation in RILS spectra from large domains of the bulk FQHE fluid and the small curvature of the magnetoroton dispersion near zero momentum. The low-lying neutral collective excitations from bulk-like FQHE fluid in large domains could be interpreted in terms of theories for uniform FQHE phases. With increasing wave vectors, dispersive long-wavelength modes observed in RILS at $v = 1/3$ split in a manner consistent with a two-roton bound state [13].

Long-wavelength magnetoroton excitations play key roles in recent geometrical theories of electron correlation in bulk FQHE phases [2,3]. Magnetoroton modes occur here as spin 2 chiral gravitons, corresponding to the long-wavelength quantum fluctuations of an internal dynamic metric [4–7]. At $v = 1/3$ only the mode with spin $S = -2$ is active in RILS experiments in the $k \rightarrow 0$ limit [4–6]. The results in Ref. 13 that reveal a magnetoroton doublet offer a key experimental insight that needs to be interpreted. In the framework of geometrical theories, a doublet may occur in RILS spectra at finite $k$ because of coupling between magnetorotons with $S = -2$ and $+2$ [4,6].

In summary, RILS and DRILS methods investigate impacts of domain textures on low-lying collective excitations in the non-uniform bulk of FQHE of 2DES in GaAs quantum structures. For domains larger than the inelastic light scattering wavelength ($\lambda = 1.2\ \mu m$) there is nearly full conservation of wave vector in the light scattering events. Magnetoroton excitations have been identified in RILS spectra from $v = 5/2$ FQHE fluids [45]. RILS measurements with higher resolution in large domains of FQHE fluids at $v = 5/2$ could identify narrow long-wavelength neutral collective excitations with well-defined wave vector dispersion. Such experiments may reveal features of topological domain textures involving Pfaffian and anti-Pfaffian orders [15–20]. Suitable design of GaAs quantum structures [28], enhanced growth protocols [52] and characterization by RILS and DRILS may lead to creation and identification of large domains of quantum fluids likely to host novel electron correlation effects.

## Acknowledgment

A. P. gratefully acknowledges illuminating discussions with F. D. M. Haldane, J. K. Jain, D. X. Nguyen, E. Rezayi and K. Yang. Z. L. gratefully acknowledges valuable discussions with D. X. Nguyen and K. Yang.
The work at Columbia University was supported by the National Science Foundation, Division of Materials Research under awards DMR-1306976 and DMR-2103965. The Alexander von Humboldt Foundation partially supported the experimental work at Columbia University. U.W. acknowledges funding by the Deutsche Forschungsgemeinschaft (DFG, German Research Foundation) under projects Wu 637/4-1, 4-2, 7-1. The work at Nanjing University was supported by the Fundamental Research Funds for the Central Universities (Grant No. 14380146) and National Natural Science Foundation of China (Grant No. 12074177). The work at Purdue University is supported by the U.S. Department of Energy grant DE-SC0020138 under the QIS Next Generation Quantum Systems program. The research at Princeton University is funded in part by the Gordon and Betty Moore Foundation's EPiQS Initiative, Grant GBMF9615 to L. N. Pfeiffer, and by the National Science Foundation MRSEC grant DMR-1420541.

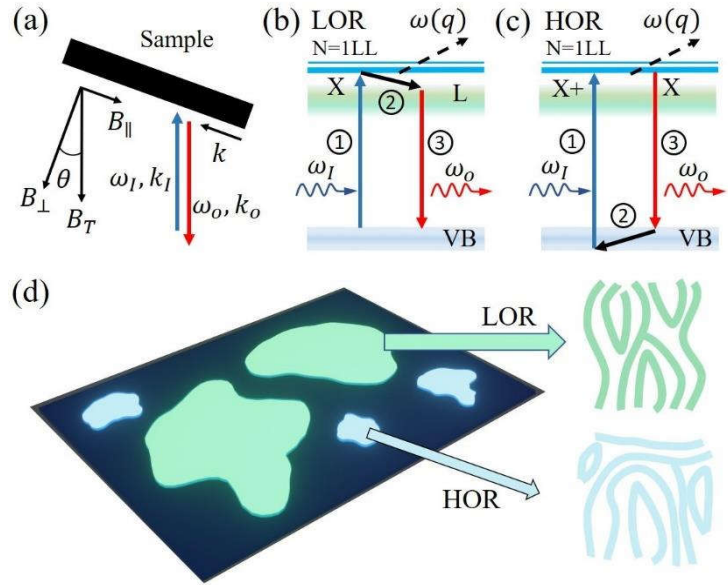

FIG. 1 (color online). (a) Schematic description of the experimental back-scattering geometry at a tilt angle $\theta = 20°$. Incoming and outgoing photons have energies $\omega_I$, $\omega_o$ and wave vectors $k_I$, $k_o$, respectively. The total magnetic field $B_T$ produces a perpendicular component $B_\perp$ and a parallel component $B_\parallel$. The inelastic scattering wave vector is parallel to $B_\parallel$. (b)(c) Different DRILS processes between valence bands and the SLL in (b) LOR and (c) HOR. A plasmon mode $\omega(q)$ is generated in the second step. (d) Schematic plot of domain textures probed by DRILS in nematic-FQHE fluids. Large (small) nematic domains are active in LOR (HOR). Smaller domains are more disordered.

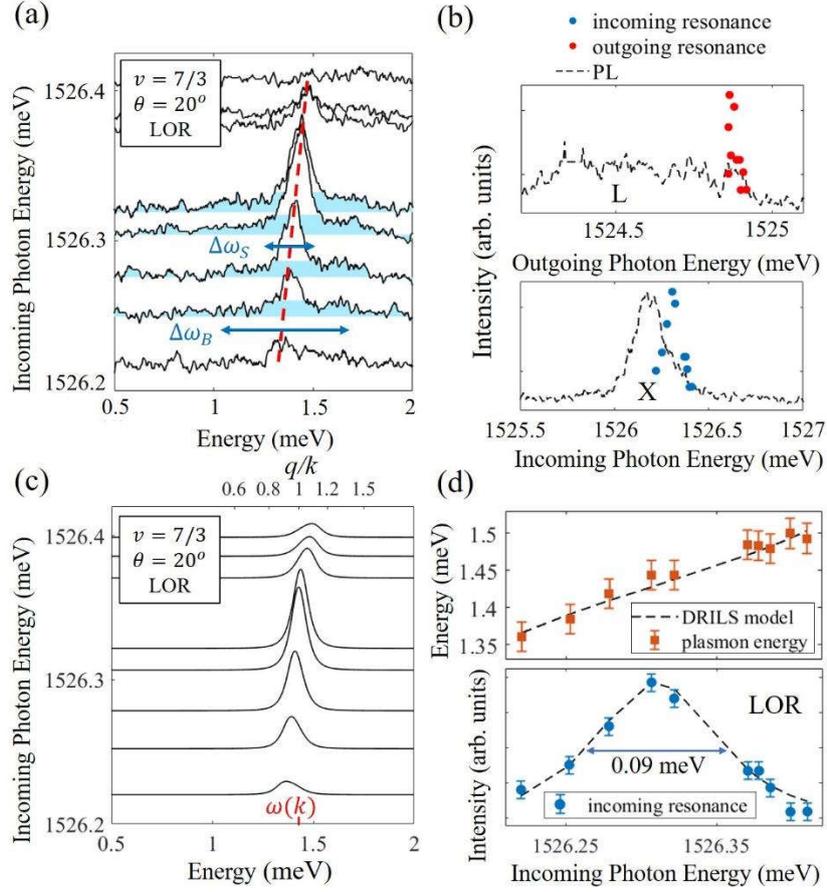

FIG. 2 (color online). (a) DRILS-LOR spectra of nematic plasmon modes measured at $v = 7/3$. The dashed line marks their blueshift. Areas filled in cyan indicate the underlying continuum. Arrows mark the different energy ranges of plasmons and the continuum. (b) Outgoing (top panel) and incoming (bottom panel) resonances of plasmons. (c) LOR plasmon spectra generated by DRILS model, where $(\omega_1, \gamma_1) = (1527.31 \text{ meV}, 0.06 \text{ meV})$, $(\omega_2, \gamma_2) = (1524.88 \text{ meV}, 0.06 \text{ meV})$, $\omega(k) = 1.43$ meV and $\Delta\omega_S = 0.16$ meV. The corresponding non-conserved wave vector $q/k$ is labelled on the top x-axis. (d) DRILS model fits (dashed lines) of plasmon energies (top panel) and incoming resonance (bottom panel) using the parameters above. The width of incoming resonance is reduced from $2\gamma_1$ due to the impact of plasmon wave vector distribution.

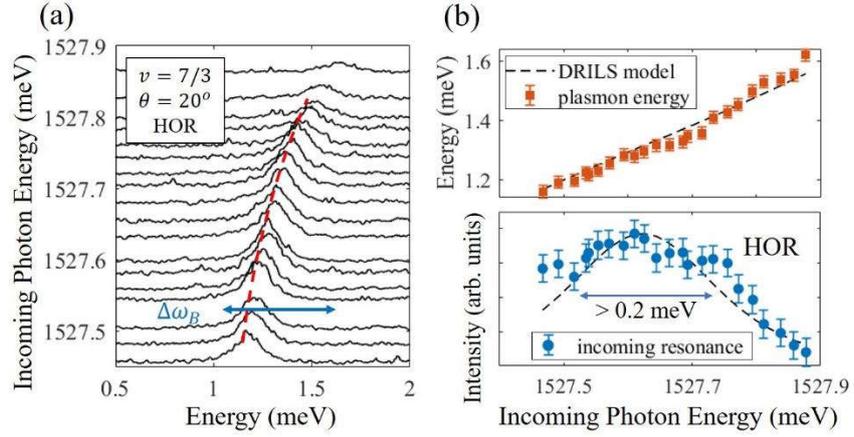

FIG. 3 (color online). (a) DRILS-HOR spectra of nematic plasmon modes within the energy range $\Delta\omega_B$ measured at $v = 7/3$. The dashed line marks their blueshift. (b) DRILS model fits (dashed lines) of plasmon energies (top panel) and incoming resonance (bottom panel) in HOR, where $(\omega_1, \gamma_1) = (1527.61$ meV, $0.25$ meV), $(\omega_2, \gamma_2) = (1526.31$ meV, $0.06$ meV), $\omega(k) = 1.32$ meV and $\Delta\omega_B = 0.35$ meV (lower limit). The incoming resonance overlaps $X+$ transitions in PL, which may be a superposition of several optical excitons.

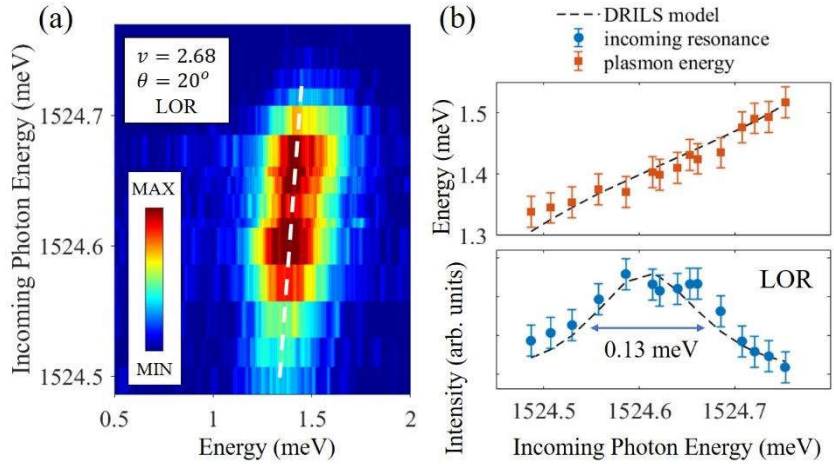

FIG. 4 (color online). (a) DRILS spectra of nematic plasmon modes at $v = 2.68$. The dashed line marks their blueshift. (b) Reproduction of plasmon energy (top panel) and incoming resonance (bottom panel) by DRILS model, where $(\omega_1, \gamma_1) = (1524.6$ meV, $0.08$ meV), $(\omega_2, \gamma_2) = (1523.21$ meV, $0.08$ meV), $\omega(k) = 1.41$ meV and $\Delta\omega = 0.24$ meV. The color code is linear with intensity.

# Supplemental Material

# Domain textures in the fractional quantum Hall effect

Ziyu Liu[1,*], Ursula Wurstbauer[2], Lingjie Du[3], Ken W. West[4], Loren N. Pfeiffer[4], Michael J. Manfra[5,6], Aron Pinczuk[1,7,‡]

[1] *Department of Physics, Columbia University, New York, New York 10027, USA*

[2] *Institute of Physics, University of Münster, Wilhelm-Klemm-Str.10, 48149 Münster, Germany*

[3] *School of Physics, and National Laboratory of Solid State Microstructures, Nanjing University, Nanjing 210093, China*

[4] *Department of Electrical Engineering, Princeton University, Princeton, New Jersey 08544, USA*

[5] *Department of Physics and Astronomy, School of Materials Engineering,*

*and School of Electrical and Computer Engineering, Purdue University, Indiana 47907, USA*

[6] *Microsoft Quantum Lab Purdue, Purdue University, West Lafayette, Indiana 47907, USA*

[7] *Department of Applied Physics and Applied Mathematics, Columbia University, New York, New York 10027, USA*

*\*zl2577@columbia.edu, ‡ap359@columbia.edu*

## 1. Optical transitions in the Second Landau Level (SLL)

Typical Photoluminescence (PL) and resonant Rayleigh scattering (RRS) spectra from the two-dimensional electron system at filling factor $v = 7/3$ are shown in Fig. S1. Three relevant optical transitions are labelled as $L$, $X$ and $X+$. Non-equilibrium $X+$ excitons are obscured by noise in PL but clearly identified from RRS. They serve as resonant channels in resonant light scattering. Resonance enhancement of inelastic light scattering intensity is achieved by tuning the incoming photon energy close to either $X$ or $X+$ excitons. In addition, when the outgoing photon energy is in the vicinity of either $L$ or $X$ transitions, double resonant inelastic light scattering (DRILS) dominates.

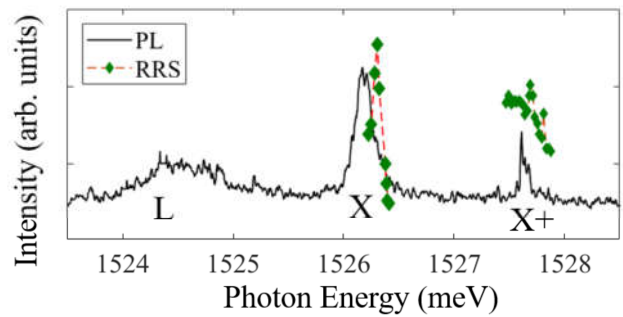

Figure S1. Photoluminescence and resonant Rayleigh scattering spectra at $v = 7/3$.

## 2. Determination of filling factors in the SLL

Filling factors $v$ of fractional quantum hall effect (FQHE) states in the SLL under illumination are determined from resonant inelastic light scattering (RILS) measurements of the long wavelength spin wave mode (SW) around $v = 3$. The SW mode is expected to show prominent enhancement when the magnetic field is tuned to define the fully spin polarized $v = 3$ state [1–3], reminiscent of the case around $v = 1$ [4]. Based on the determination of magnetic field strength at $v = 3$, magnetic fields for other filling factors in the SLL can be calculated.

RILS measurements of SW around $v = 3$ are shown in Fig. S2(a). A slight deviation of magnetic field from that of $v = 3$ significantly reduces the mode intensity of fully enhanced RILS spectra. Spectra at extreme resonances are plotted in Fig. S2(b). The SW modes at the Zeeman energy show a large enhancement precisely at $v = 3$. The determined field strength $B = 3.8$ T (total field with a tilt between sample normal and magnetic field of about $\theta = 20°$) for $v = 3$ yields the electron density under resonant illumination at 45 mK to be $2.8 \times 10^{15}$ m$^{-2}$, close to the value ($2.9 \times 10^{15}$ m$^{-2}$) measured by transport experiments at low temperature done on a different piece from the same wafer without an in-plane magnetic field component.

The RILS intensity ratio between Stokes and anti-Stokes modes is expected to be $e^{\hbar\omega/k_B T}$. A resistance thermometer is used to measure the temperature and gives an upper bound around 45 mK. For SW with energy around 0.1 meV, the measured temperature around 45 mK is consistent with the vanishing of anti-Stokes modes shown in Fig. S2. We note that both electron temperature and density are stable against the wavelength or power of incoming photons (over several decades of excitation intensities), confirmed by fitting the Fermi edge in PL measurements to the Fermi-Dirac distribution at zero magnetic field. Moreover, the line-width of collective intersubband excitation (charge density excitation) as a signature for purity of the system is independent of the above external illumination conditions.

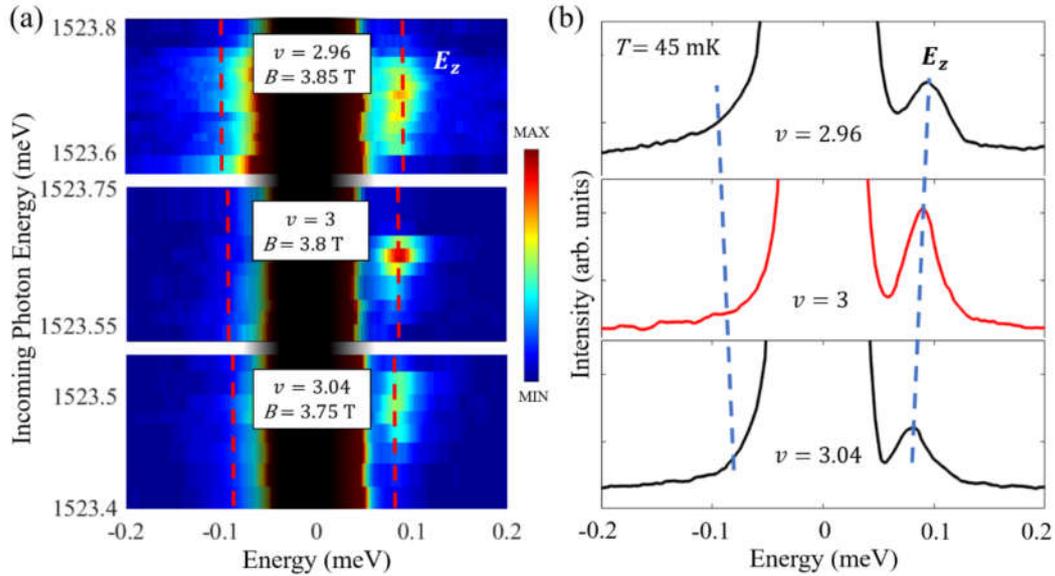

Figure S2. (a) RILS of SW at $v = 2.96$, 3 and 3.04. Rayleigh scattering signals are shaded. Dashed lines mark the Zeeman energies at each filling factor. The color code is linear with intensity. (b) Spectra at maximum resonance in (a) show prominent enhancement of SW at $v = 3$. The SW energies shift along Zeeman energies (dashed line), and only Stokes modes are observed.

### 3. Well-defined FQHE state at $v = 7/3$

The existence of a well-defined FQHE state is the prerequisite for investigating impacts of domain textures in FQHE-nematic fluids under tilted magnetic field. Figure S3 presents temperature and filling factor dependence of magnetoroton excitations $E_g$ around $v = 7/3$. Rapid suppression of magnetoroton excitations at ν = 7/3 with increasing temperatures or by tuning the applied magnetic field slightly away from ν = 7/3 confirms that FQHE is well defined at ν = 7/3. Similar behavior of magnetoroton excitations at other filling factors in the SLL are reported in previous works [2,3].

The observation that both magnetoroton excitations and plasmons are excited under a similar resonant condition at ν = 7/3 is key evidence of the phase coexistence in FQHE-nematic fluids. Their evolution as a function of in-plane magnetic field in the SLL has not been fully resolved. Transport measurements show that with increasing in-plane magnetic field (tilt angle), FQHE state may collapse [5,6], strengthen [6,7] or form a novel nematic FQHE state [8] which may be related to heterostructure design parameters. A detailed optical investigation probing the bulk of the FQHE-nematic fluids is of great interests but is beyond the scope of the current work.

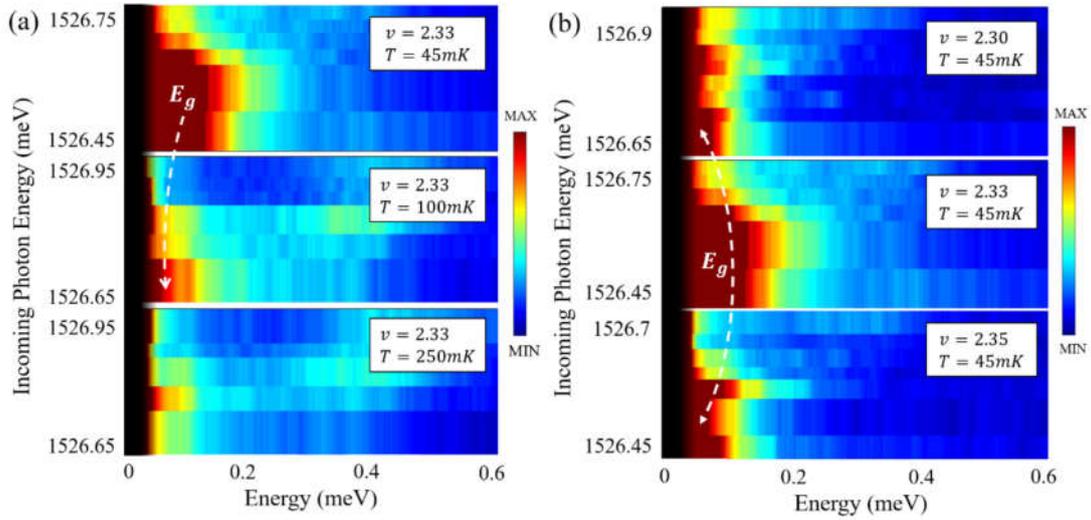

Figure S3. (a) Suppression of magnetoroton excitations with increasing temperatures at ν = 7/3. (b) Suppression of magnetoroton excitations by detuning the filling factor ν away from ν = 7/3 at 45 mK. The color code is linear with intensity.

## 4. DRILS intensity model

We set up a model describing DRILS by plasmons based on 3$^{rd}$ order time dependent perturbation theory [9]. For a uniform two-dimensional electron system (2DES) in general, the light scattering intensity is proportional to the square of matrix element

$$I \propto \left| \sum_{m,n} \frac{H_{fn} P_{(n,q,k_o)(m,k_I)} H_{mi}}{(\omega_I - \omega_m - i\gamma_m)(\omega_o - \omega_n - i\gamma_n)} \right|^2 \delta(\boldsymbol{k_I} - \boldsymbol{k_o} - \boldsymbol{q})$$

where $H$ represents light-matter coupling and $P$ stands for plasmon-electron interactions. $(m,n)$ are the intermediate states of the system with energies $(\omega_m, \omega_n)$ and broadenings $(\gamma_m, \gamma_n)$. The denominator is a double resonance term where $(m,n)$ states serve as incoming and outgoing channels resonating with incoming $(\omega_I)$ and outgoing $(\omega_o)$ photons, respectively. The delta function represents wave vector conservation between incoming photon $(\boldsymbol{k_I})$, outgoing photon $(\boldsymbol{k_o})$ and plasmon (***q***).

We assume that the energy dependences of *H* and *P* only play a minor role compared to the double resonance term. Only one pair of incoming and outgoing channels (*m,n*) are considered in each resonant range. The DRILS intensity then becomes

$$I \propto \left|\frac{1}{(\omega_I-\omega_1-i\gamma_1)}\frac{1}{(\omega_o-\omega_2-i\gamma_2)}\right|^2 \delta(\boldsymbol{k_I} - \boldsymbol{k_o} - \boldsymbol{q})$$

In non-uniform electron fluids, wave vector conservation is broken under a finite domain size *D*. The delta function can be replaced with a Lorentzian to account for the resultant finite wave vector uncertainty $\Delta q = 2\pi/D$ [6]. We further write the Lorentzian as a function of energy following the plasmon dispersion $\omega(q)$

$$\delta(\boldsymbol{k_I} - \boldsymbol{k_o} - \boldsymbol{q}) \to \frac{1}{(q-k)^2+(\Delta q/2)^2} \propto \frac{1}{(\omega(q)-\omega(k))^2+(\Delta\omega/2)^2}$$

$$\omega(q) = (1+\xi)\sqrt{n^* e^2 q/(2\varepsilon\varepsilon_0 m^*)}$$

$k = |\boldsymbol{k_I} - \boldsymbol{k_o}|sin\theta$ is the finite wave vector transfer in back-scattering geometry. $\omega(q)$ is valid in the limit $\lambda \gg l_0$, where $\lambda$ is the inelastic scattering wavelength, $l_0$ is the magnetic length, and the factor $\xi$ accounts for the difference with a uniform 2DES [10]. $m^*$ is the band mass of 2DES in the GaAs quantum well. $\varepsilon$ and $\varepsilon_0$ are the background dielectric constant and free-space permittivity, respectively. $n^*$ is the quasiparticle density in the spin-up SLL. We finally obtain Eq. (1) in main text

$$I \propto \left|\frac{1}{(\omega_I-\omega_1-i\gamma_1)}\frac{1}{(\omega_o-\omega_2-i\gamma_2)}\right|^2 \frac{1}{(\omega(q)-\omega(k))^2+(\Delta\omega/2)^2} \tag{S1}$$

To observe DRILS signals from certain collective excitations, a pair of excited quantum states of the system is needed to serve as incoming and outgoing resonant channels. Moreover, their energy difference should match the energy of the low-lying collective modes. These conditions are not necessarily satisfied by magnetoroton modes in the FQHE. For magnetorotons in the SLL, they have much smaller energies (~ 0.1 meV) compared to plasmons which may induce mixing of incoming and outgoing channels. For magnetorotons in the lowest Landau level, a series of charge- and spin- excitations with different resonance conditions could complicate the interpretations [11].

## 5. Quantitative analysis of the DRILS intensity

In Eq. (S1), energies and broadenings of resonant channels in each set of spectra are determined from RRS data, while $\omega(k)$ and $\Delta\omega$ are uncertain. The domain size is related to the ratio between $\omega(k)$ and $\Delta\omega$. To determine the parameters which best reproduce the measured DRILS spectra, we numerically generate DRILS spectra under different values of $\omega(k)$ and $\Delta\omega$, and evaluate the error (sum of squared residuals) in incoming resonance and plasmon energies.

The color map of the error provides estimates of $\omega(k)$ and $\Delta\omega$, thus domain sizes. Figure S4(a) shows the error map for DRILS-LOR, and gives a bounded parameter space. In contrast, for DRILS-HOR shown in Fig. S4(b), $\omega(k)$ and $\Delta\omega$ are partially bounded. As a result, the model is only able to give an upper bound of the domain sizes in HOR. DRILS model thus better accounts for LOR spectra.

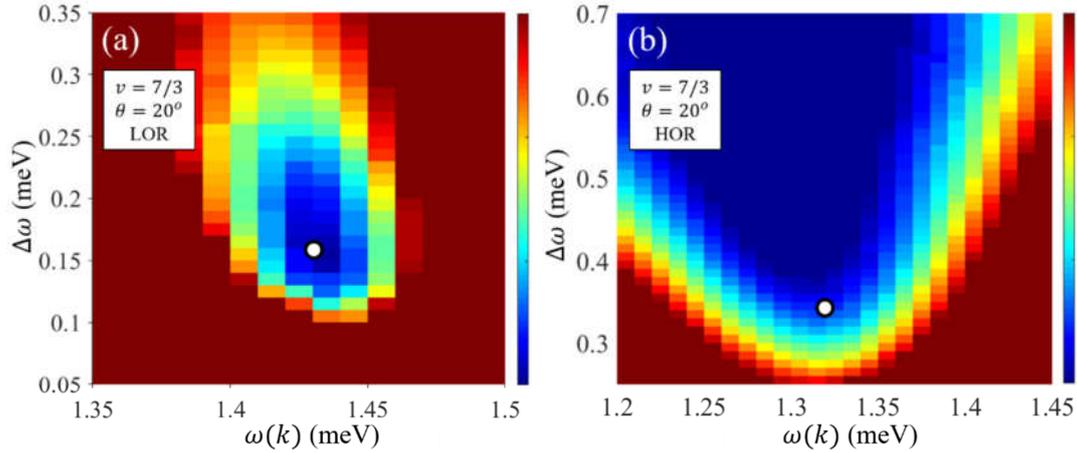

Figure S4. (a) Map of the modelling quality at $v = 7/3$ in LOR. The white dot marks the parameters of fit results presented in Fig. 2. (b) Map of the modelling quality at $v = 7/3$ in HOR. The white dot marks the parameters of fit results presented in Fig. 3. The color code is linear with intensity.